\documentstyle[preprint,aps]{revtex}

\begin{document}

\draft

\title{
Mirror matter admixtures in $K_{\rm L}\to\gamma\gamma$
}

\author{
G.~S\'anchez-Col\'on
}
\address{
Departamento de F\'{\i}sica Aplicada.\\
Centro de Investigaci\'on y de Estudios Avanzados del IPN. Unidad Merida.\\
A.P. 73, Cordemex. M\'erida, Yucat\'an, 97310. MEXICO.
}
\author{
A.~Garc\'{\i}a
}
\address{
Departamento de F\'{\i}sica.\\
Centro de Investigaci\'on y de Estudios Avanzados del IPN.\\
A.P. 14-740. M\'exico, D.F., 07000. MEXICO.
}

\date{\today}

\maketitle

\begin{abstract}
Based on possible, albeit tiny, admixtures of mirror matter in ordinary
mesons we study the $K_{\rm L}\to\gamma\gamma$ transition. We find
that this process can be described with a small $SU(3)$ symmetry
breaking of only $3\%$. We also determine the $\eta$-$\eta'$ mixing angle
and the pseudoscalar decay constants. The results for these
parameters are consistent with some obtained in the literature. They
favor two recent determinations; one based on two analytical
constraints, and another one based on next-to-leading order power
corrections.
\end{abstract}

\pacs{
PACS number(s):
12.90.+b, 13.20.Eb, 14.40.Aq
}

The rare decay process $K_{\rm L}\to\gamma\gamma$ is a flavour-changing
radiative transition, which is expected to be of the same order in weak and
electromagnetic couplings as $K_{\rm L}\to\mu^+\mu^-$, yet the branching
ratio of the former is $(5.96\pm 0.15)\times 10^{-4}$, whereas that of the
latter is only $(7.25\pm 0.16)\times 10^{-9}$. The strong suppression of
$K_{\rm L}\to\mu^+\mu^-$ versus $K^+\to\mu^+\nu$, for example, is
understood as being due to the Glashow-Iliopoulos-Maiani mechanism, but
$K_{\rm L}\to\gamma\gamma$ does not appear to be suppressed at all by
this mechanism. The process $K_{\rm L}\to\gamma\gamma$ is likely to be
dominated by low-energy contributions~\cite{ma}.

In the present work we use a phenomenological model based on parity and
flavour admixtures of mirror matter in ordinary mesons~\cite{apriori},
where the $K_{\rm L}\to\gamma\gamma$ amplitude is assumed to be enhanced
by parity and flavour conserving amplitudes,
$\pi^0\to\gamma\gamma$ and $\eta_8\to\gamma\gamma$, arising with such
admixtures via the ordinary electromagnetic interaction Hamiltonian as the
transition operator. With experimental inputs and previous results for the
mixing angles of the mirror matter admixtures, we determine the
$\eta$-$\eta'$ mixing angle $\theta$ and the pseudoscalar decay
constants ratios $f_8/f_{\pi}$ and $f_1/f_{\pi}$.

In a model with mirror matter mixings, the physical mesons $K^0_{\rm ph}$
and $\bar{K}^0_{\rm ph}$ with parity and $SU(3)$-flavor violating
admixtures are given by~\cite{kpipi}
 
\[
K^0_{\rm ph} =
K^0_{\rm p} -
\frac{1}{\sqrt 2} \sigma \pi^0_{\rm p} +
\sqrt{\frac{3}{2}} \sigma \eta_{\rm 8\,p} +
\sqrt{\frac{2}{3}} \delta \eta_{\rm 8\,s} -
\frac{1}{\sqrt 3} \delta \eta_{\rm 1\,s} -
\frac{1}{\sqrt 2} \delta' \pi^0_{\rm s} +
\frac{1}{\sqrt 6} \delta' \eta_{\rm 8\,s} +
\frac{1}{\sqrt 3} \delta' \eta_{\rm 1\,s},
\]

\begin{equation}
\bar{K}^0_{\rm ph} =
\bar{K}^0_{\rm p} -
\frac{1}{\sqrt{2}} \sigma \pi^0_{\rm p} +
\sqrt{\frac{3}{2}} \sigma \eta_{\rm 8\,p} -
\sqrt{\frac{2}{3}} \delta \eta_{\rm 8\,s} +
\frac{1}{\sqrt{3}} \delta \eta_{\rm 1\,s} +
\frac{1}{\sqrt{2}} \delta' \pi^0_{\rm s} -
\frac{1}{\sqrt{6}} \delta' \eta_{\rm 8\,s} -
\frac{1}{\sqrt{3}} \delta' \eta_{\rm 1\,s}.
\label{uno}
\end{equation}

\noindent
We have used the $SU(3)$-phase conventions of Ref.\cite{deswart}. The
mixing angles $\sigma$, $\delta$, and $\delta'$, are the parameters of the
model, which have been fitted previously~\cite{detailed,universality}; see
later on. The subindeces ${\rm s}$ and ${\rm p}$ refer to positive and
negative parity eigenstates, respectively.  Notice that the physical
mesons satisfy $CPK^0_{\rm ph}=-\bar{K}^0_{\rm ph}$ and $CP\bar{K}^0_{\rm
ph}=-K^0_{\rm ph}$.

We can form the $CP$-eigenstates $K_{\rm 1}$ and $K_{\rm 2}$ as

\begin{equation}
K_{\rm 1_{ph}} = \frac{1}{\sqrt{2}} (K^0_{\rm ph} - \bar{K}^0_{\rm ph})
\qquad\mbox{and}\qquad
K_{\rm 2_{ph}} = \frac{1}{\sqrt{2}} (K^0_{\rm ph} + \bar{K}^0_{\rm ph}),
\label{dos}
\end{equation}

\noindent
the $K_{\rm 1_{ph}}$ ($K_{\rm 2_{ph}}$) is an even (odd) state with respect
to $CP$. Here, we shall not consider $CP$-violation and therefore,
$|K_{\rm S,L}\rangle = |K_{1,2}\rangle$.

Substituting the expressions given in Eqs.~(\ref{uno}), we obtain,

\[
K_{\rm S_{ph}} =
K_{\rm S_p} +
\frac{1}{\sqrt{3}} (2\delta + \delta') \eta_{\rm 8\,s} -
\delta' \pi^0_{\rm s} -
\sqrt{\frac{2}{3}} (\delta - \delta') \eta_{\rm 1\,s},
\]

\begin{equation}
K_{\rm L_{ph}} =
K_{\rm L_p} -
\sigma \pi^0_{\rm p} +
\sqrt{3} \sigma \eta_{\rm 8\,p},
\label{tres}
\end{equation}

\noindent
where the usual definitions
$K_{\rm 1_p} = (K^0_{\rm p} - \bar{K}^0_{\rm p})/\sqrt{2}$
and
$K_{\rm 2_p} = (K^0_{\rm p} + \bar{K}^0_{\rm p})/\sqrt{2}$
were used.

Mirror matter admixtures in the physical mesons can contribute to the
$K_{\rm S,L}\to\gamma\gamma$ amplitudes via the ordinary parity and
flavour-conserving electromagnetic interaction Hamiltonian $H^{\rm em}$.
Using Eqs.~(\ref{tres}), a very simple calculation leads to

\begin{equation}
F_{K_{\rm S}\gamma\gamma} =
\frac{1}{\sqrt{3}}(2\delta+\delta')F_{\eta_{\rm 8\,s}\gamma\gamma} -
\delta' F_{\pi^0_{\rm s}\gamma\gamma} -
\sqrt{\frac{2}{3}}(\delta-\delta')F_{\eta_{\rm 1\,s}\gamma\gamma},
\label{cuatro}
\end{equation}

\begin{equation}
F_{K_{\rm L}\gamma\gamma} =
-\sigma F_{\pi^0_{\rm p}\gamma\gamma} +
\sqrt{3}\sigma F_{\eta_{\rm 8\,p}\gamma\gamma},
\label{cinco}
\end{equation}

\noindent
where
$F_{K_{\rm S}\gamma\gamma}=
\langle\gamma\gamma|H^{\rm em}|K_{\rm S_{ph}}\rangle$,
$F_{K_{\rm L}\gamma\gamma}=
\langle\gamma\gamma|H^{\rm em}|K_{\rm L_{ph}}\rangle$,
and
$F_{\pi^0_{\rm s}\gamma\gamma}=\langle\gamma\gamma
|H^{\rm em}|\pi^0_{\rm s}\rangle$, etc.

Given that $K_{\rm S}$ and $K_{\rm L}$ are $CP=+1$ and $CP=-1$ pure states
respectively, and because the two-photon state is a $C=+1$ state, then
$K_{\rm S}\to\gamma\gamma$ must go through a so-called parity-violating
transition while $K_{\rm L}\to\gamma\gamma$ goes through a
parity-conserving transition. In the first case the two-photon final state
is $P=+1$ while in the second one, $P=-1$. However, as we can see from
Eqs.~(\ref{cuatro}) and (\ref{cinco}), in the context of mirror matter
admixtures all the contributions to both amplitudes are flavour and parity
conserving. Notice that the additive terms on the right-hand side
of these equations involve only mirror mesons in $F_{K_{\rm
S}\gamma\gamma}$ and only ordinary mesons in $F_{K_{\rm L}\gamma\gamma}$.

We can see from (\ref{cinco}) that the parity-conserving amplitude
$F_{K_{\rm L}\gamma\gamma}$ vanish in the  strong-flavour $SU(3)$-symmetry
limit ($U$-spin invariance):

\begin{equation}
F_{\eta_8\gamma\gamma}=\frac{1}{\sqrt{3}}F_{\pi^0\gamma\gamma}.
\label{cinco1}
\end{equation}

\noindent
The so-called parity-violating amplitude $F_{K_{\rm S}\gamma\gamma}$
remains non-zero in this limit. This is the same result obtained as a
theorem previously~\cite{limit}.

Let us now concentrate on $K_{\rm L}\to\gamma\gamma$. As a first
approximation, we shall compare Eq.~(\ref{cinco}) directly with experiment
by ignoring any other existent contributions. The two-photon decay widths
for $P^0=K_{\rm L}, \pi^0, \eta, \eta'$ can be expressed as

\begin{equation}
\Gamma(P^0\to\gamma\gamma) =
\frac{F^2_{P^0\gamma\gamma}m^3_{P^0}}{64\pi},
\label{seis}
\end{equation}

\noindent
with the decay amplitudes given by the matrix elements
$F_{P^0\gamma\gamma} = \langle\gamma\gamma|H|P^0\rangle$,
with $H=H^{\rm em}$ as the transition operator. From the present experimental
values~\cite{pdg} of the decay rates we determine the observed values for
the $2\gamma$-amplitudes, they are displayed in Table~\ref{table1}. Of the
values for the mixing angles of the mirror matter admixtures obtained
previously~\cite{detailed,universality}, we shall only need
$\sigma=(4.9\pm2.0)\times10^{-6}$. We do not quote the values of the other
two mixing angles because we shall not use them here.

The parametrization of the $\eta$-$\eta'$ mixing has been the subject of
many studies for already many years, as can be appreciated in the
corresponding reviews in Refs.~\cite{pdg} and \cite{feldmannpdg}. Based on
theoretical arguments and on detailed phenomenological analyses, it has
been generally accepted that the $\eta$-$\eta'$ mixing cannot be described
in a process independent fashion by applying the same rotation (with one
angle only) simultaneously to octet-singlet states and to their decay
constants. Two separate rotations should be used and two mixing angles are
required. In our analysis we shall follow the prescription discussed by Cao
and Signal~\cite{cao}, which allows one to still use one-mixing angle at
the state level. That is, we shall introduce this angle at the amplitude
level only and shall not make the questionable assumption that it
also applies to the mixing of the decay constants. Accordingly, we can
write at the amplitude level

\begin{equation}
F_{\eta_8\gamma\gamma} = F_{\eta'\gamma\gamma}\sin\theta +
F_{\eta\gamma\gamma}\cos\theta,
\label{nueve}
\end{equation}

\begin{equation}
F_{\eta_1\gamma\gamma}=F_{\eta'\gamma\gamma}\cos\theta -
F_{\eta\gamma\gamma}\sin\theta,
\label{ocho}
\end{equation}

\noindent
and the pseudoscalar decay constants ratios are given by

\begin{equation}
F_{\eta_8\gamma\gamma}=\frac{\alpha}{\pi}
\frac{1}{\sqrt{3}}\frac{1}{f_{\pi}}\left(\frac{f_{\pi}}{f_8}\right)
\qquad {\rm and} \qquad
F_{\eta_1\gamma\gamma}=\frac{\alpha}{\pi}
\frac{2\sqrt{2}}{\sqrt{3}}\frac{1}{f_{\pi}}\left(\frac{f_{\pi}}{f_1}\right),
\label{diez}
\end{equation}

\noindent
with $\sqrt{2}f _{\pi}=(130.7\pm0.3){\rm MeV}$~\cite{pdg}.

Before proceeding further, let us make a first estimation. In the previous
studies the predictions for $\theta$ vary from
$-10^{\circ}$~\cite{isgur} to $-23^{\circ}$~\cite{donoghue,gilman}, the
ones for $f_8$ from $(0.94)f_{\pi}$~\cite{cao} to
$(1.38)f_{\pi}$~\cite{venugopal}, and the ones for $f_1$ from
$(1.04)f_{\pi}$~\cite{donoghue,gilman} to $(1.17)f_{\pi}$~\cite{cao}.
Within the chiral anomaly sector the most commonly accepted values for the
mixing parameters are~\cite{venugopal}: $\theta\approx-20^{\circ}$,
$f_8/f_{\pi}\approx1.3$, and $f_1/f_{\pi}\approx1.0$. Let us use these
values in Eq.~(\ref{nueve}) First, notice that we can not determine the
signs of the $2\gamma$-amplitudes from the decays widths so, for
definiteness, we will choose $F_{\pi^0\gamma\gamma}$ as positive. In this
case, $F_{\eta_8\gamma\gamma}$ is also positive, as we can see from the
$SU(3)$-symmetry limit relation. Besides, for this value of $\theta$ we
find from Table~\ref{table1} that to get the best agreement with
(\ref{cinco1}), the phases of $F_{\eta\gamma\gamma}$ and
$F_{\eta'\gamma\gamma}$ have to be set as positive too. Then, from
(\ref{nueve}) we obtain $F_{\eta_8\gamma\gamma}=1.174\times 10^{-5}\ {\rm
MeV}^{-1}$, with an $SU(3)$-symmetry breaking of $19\%$. Finally, from
expression~(\ref{cinco}) for $F_{K_{\rm L}\gamma\gamma}$ in the context of
mirror matter admixtures, and the observed values of $F_{\pi^0\gamma\gamma}$ and
$\sigma$, we find $F_{K_{\rm L}\gamma\gamma}=-0.236\times 10^{-12}\
{\rm MeV}^{-1}$, to be compared with $|F_{K_{\rm
L}\gamma\gamma}|= 3.519\times 10^{-12}\ {\rm MeV}^{-1}$. So, we are
an order of magnitude down from experiment by using the predictions
of chiral perturbation theory (ChPT) for the $\eta$-$\eta'$ parameters.
Thus, $K_{\rm L}\to\gamma\gamma$ is quite sensitive to such mixing.

We can now proceed to determine the value of $\theta$ from the
observed values for the $2\gamma$-amplitudes of Table~\ref{table1} and
$\sigma$. We shall preserve though, the positive phases for the decay
amplitudes of  $\pi^0,\eta,\eta'\to\gamma\gamma$. The phase of
the decay amplitude for the rare transition $K_{\rm L}\to\gamma\gamma$,
to be used in the mirror matter admixtures relation
(\ref{cinco}), will remain free. From this relation we obtain now,
$F_{\eta_8\gamma\gamma}= (1.494\pm 0.054)\times 10^{-5}\ {\rm MeV}^{-1}$
if $F_{K_{\rm L}\gamma\gamma}>0$ and $F_{\eta_8\gamma\gamma}=
(1.411\pm 0.054)\times 10^{-5}\ {\rm MeV}^{-1}$ if $F_{K_{\rm
L}\gamma\gamma}<0$. In this case there is only a $\pm2.9\%$ $SU(3)$-symmetry
breaking, respectively. In other words, the $K_{\rm L}\to\gamma\gamma$
process is described in the mirror matter admixtures context (relation
(\ref{cinco}) and the independently determined value of $\sigma$) with
just a small $SU(3)$-symmetry flavour breaking of $2.9\%$. This is made clear if
we parametrize the violation of the $SU(3)$-relation (\ref{cinco1}) as

\begin{equation}
F_{\eta_8\gamma\gamma}=\frac{1}{\sqrt{3}}F_{\pi^0\gamma\gamma}(1+b_3),
\label{cinco2}
\end{equation}

\noindent
so that (\ref{cinco}) transforms into

\begin{eqnarray}
F_{K_{\rm L}\gamma\gamma} &=&
\sigma F_{\pi^0_{\rm p}\gamma\gamma}b_3
\nonumber\\
&\approx&
(4.9\times10^{-6})(2.5\times10^{-5}{\rm MeV}^{-1})(2.9\times10^{-2})
\nonumber\\
&\approx&
3.5\times10^{-12}{\rm MeV}^{-1},
\label{cinco3}
\end{eqnarray}

\noindent
as experimentally required.

The corresponding value for the $\eta$-$\eta'$ mixing angle is now
determined from Eq.~(\ref{nueve}) and the pseudoscalar decay constants
ratios $f_1/f_{\pi}$ and $f_8/f_{\pi}$ are obtained using Eqs.~(\ref{ocho})
and (\ref{diez}). The results of this approach are shown in row~I of
Table~\ref{table2}, where for the sake of comparison we have included the
results of previous determinations of the $\eta$-$\eta'$ mixing
parameters.

A different possibility in our analysis is revealed by noticing that in the
context of mirror matter admixtures for physical hadrons, all the
amplitudes on the right hand side of Eqs.~(\ref{cuatro}) and (\ref{cinco})
may be affected by the mass of the physical $K^0_{ph}$ involved in the
decaying $K_{\rm S_{ph}}$ and $K_{\rm L_{ph}}$, respectively. At this
point, we have ignored such dependence. We will take this into account by
changing the normalization of  $F_{\pi\gamma\gamma}$,
$F_{\eta_8\gamma\gamma}$, and $F_{\eta_1\gamma\gamma}$ in Eqs.~(\ref{cuatro})
and (\ref{cinco}), from $f_{\pi}$ to $f_K$, with $\sqrt{2}f_K=(159.8\pm1.6){\rm
MeV}$~\cite{pdg}. This means,

\begin{equation}
F_{\pi\gamma\gamma}=\frac{\alpha}{\pi}
\frac{1}{f_{\pi}}
\to
\frac{\alpha}{\pi}
\frac{1}{f_K}\equiv
F^{(K)}_{\pi\gamma\gamma},
\label{once}
\end{equation}

\begin{equation}
F_{\eta_8\gamma\gamma}=\frac{\alpha}{\pi}
\frac{1}{\sqrt{3}}\frac{1}{f_{\pi}}\left(\frac{f_{\pi}}{f_8}\right)
\to
\frac{\alpha}{\pi}
\frac{1}{\sqrt{3}}\frac{1}{f_K}\left(\frac{f_{\pi}}{f_8}\right)\equiv
F^{(K)}_{\eta_8\gamma\gamma},
\label{doce}
\end{equation}

\begin{equation}
F_{\eta_1\gamma\gamma}=\frac{\alpha}{\pi}
\frac{2\sqrt{2}}{\sqrt{3}}\frac{1}{f_{\pi}}\left(\frac{f_{\pi}}{f_1}\right)
\to
\frac{\alpha}{\pi}
\frac{2\sqrt{2}}{\sqrt{3}}\frac{1}{f_K}\left(\frac{f_{\pi}}{f_1}\right)\equiv
F^{(K)}_{\eta_1\gamma\gamma}.
\label{trece}
\end{equation}

\noindent
From (\ref{once}) we find the value for the intermediate transition
$\pi\to\gamma\gamma$ normalized to $f_K$:
$F^{(K)}_{\pi\gamma\gamma}=(2.056\pm0.020)\times10^{-5}\ {\rm MeV}^{-1}$.
With this value and repeating the steps of the previous analysis we obtain,
from the mirror matter admixtures relation (\ref{cinco}),
$F^{(K)}_{\eta_8\gamma\gamma}= (1.229\pm 0.021)\times 10^{-5}\ {\rm
MeV}^{-1}$ if $F_{K_{\rm L}\gamma\gamma}>0$ and
$F^{(K)}_{\eta_8\gamma\gamma}= (1.145\pm 0.021)\times 10^{-5}\ {\rm
MeV}^{-1}$ if $F_{K_{\rm L}\gamma\gamma}<0$,
with a $\pm3.5\%$ $SU(3)$-symmetry breaking, respectively.
As above, from Table~\ref{table1} and Eq.~(\ref{nueve}) we determine
$\theta$, while $f_8/f_{\pi}$ is evaluated using now
(\ref{doce}) and $f_1/f_{\pi}$ is determined from (\ref{ocho}) and
(\ref{trece}). The results of this approach are displayed in row~II of
Table~\ref{table2}.

As we can see from Table~\ref{table2}, our predictions for the
$\eta$-$\eta'$ mixing angle are consistent with those reported in the
literature. The results of row~I agree with those given in
Refs.~\cite{cao,crystal,feldmann,escribano,yeh,crystal92,bramon}, which
were obtained by considering various decay processes. The $\theta$-values
of row~II are consistent with the predictions based on the chiral
Lagrangian and phenomenological mass
formulas~\cite{donoghue,gilman,venugopal,burakovsky,choi}. Also, our
results for $f_8/f_{\pi}$ are smaller than the most accepted prediction of
ChPT, $f_8/f_{\pi}=1.3$, and most phenomenological analyses, but they are
in agreement with the values obtained in Refs.~\cite{cao,yeh}. Our
predictions for the ratio $f_1/f_{\pi}$ of row~I are consistent with all
the previous determinations reported, but those of row~II are smaller than
them.

In summary, in the framework of mirror matter admixtures, the
description of the $K_{\rm L}\to\gamma\gamma$ process is possible
and only requires an $SU(3)$-symmetry breaking of just a $3\%$. Also, the
results for the $\eta$-$\eta'$ mixing parameters in this approach are
consistent with the values obtained in the literature. Two important
remarks are in order here. First, the approach and the data used here and
in references~\cite{cao} and \cite{yeh} are very different. Ref.~\cite{cao}
obtains two analytical constraints on the parameters by considering the
two-photon decays of $\eta$ and $\eta'$, and the production of these
states in $e^+e^-$ scattering at large momentum transfer, the parameters
are determined from the data on the decay processes and CLEO measurements
on the meson-photon transition form factors. Ref~\cite{yeh} evaluates the
next to leading order power corrections to the $\eta\gamma$ and
$\eta'\gamma$ form factors and employs them to evaluate the $\eta$-$\eta'$
mixing parameters. Second, the agreement of our results in row~I and those
of these two references lends support to the expectation that the
parametrization of this mixing phenomenon is both process and energy
independent, an expectation that has been clearly emphasized by
Feldmann~\cite{feldmannpdg}. It is in this spirit that the results of row~I
are more attractive than those of row~II. Finally, let us stress once more
that our one-angle mixing scheme does not use the questionable assumption
of applying the same rotation to the decay constants and, as emphasized in
Ref.~\cite{cao}, our favored values must be taken as suggestive, too. A
connection with other parametrizations deserves further study and we hope
to address it in the near future.

We would like to thank CONACyT (M\'exico) for partial support.

\begin{table}
\caption{
Observed values for the $2\gamma$ decay amplitudes of $K_{\rm L}$, $\pi^0$,
$\eta$, and $\eta'$ in ${\rm MeV}^{-1}$.
}
\label{table1}
\begin{tabular}
{
l
r@{$\,\pm\,$}l
}
$|F_{K_{\rm L}\gamma\gamma}|$ & $(3.519$ &
$0.046)\times10^{-12}$ \\
$|F_{\pi^0\gamma\gamma}|$ & $(2.516$ &
$0.089)\times10^{-5}$ \\
$|F_{\eta\gamma\gamma}|$ & $(2.39$ &
$0.11)\times10^{-5}$ \\
$|F_{\eta'\gamma\gamma}|$ & $(3.13$ &
$0.16)\times10^{-5}$ \\
\end{tabular}
\end{table}

\begin{table}
\caption{
Comparison of different determinations of the mixing parameters $\theta$,
$f_8/f_{\pi}$, and $f_1/f_{\pi}$. The values of rows~I and II were
determined in the present work. Row~I  by direct comparison with
experiment and row~II by normalizing the decay amplitudes to the $f_K$
decay constant. Upper and lower values in these rows correspond to
the choices $F_{K_{\rm L}\gamma\gamma}>0$ and $F_{K_{\rm
L}\gamma\gamma}<0$, respectively.
}
\label{table2}
\begin{tabular}
{
l
l@{$\,\pm\,$}l
l@{$\,\pm\,$}l
l@{$\,\pm\,$}l
}
Ref. &
\multicolumn{2}{c}{$\theta^{\circ}$} &
\multicolumn{2}{c}{$f_8/f_{\pi}$} &
\multicolumn{2}{c}{$f_1/f_{\pi}$} \\
\tableline
\ \ I &
$-$15.1 & 1.6 &
0.971 & 0.035 & 1.128 & 0.056 \\
&
$-$16.4 & 1.6&
1.028 & 0.039 & 1.115 & 0.055 \\
\ \ II &
$-$19.2 & 1.4 &
0.966 & 0.019 & 0.895 & 0.039 \\
&
$-$20.5 & 1.4 &
1.036 & 0.022 & 0.890 & 0.039 \\
\cite{donoghue,gilman} &
$-$23 & 3 &
\multicolumn{2}{l}{1.25} &
1.04 & 0.04 \\
\cite{cao} &
$-$14.5 & 2.0 &
0.94 & 0.07 & 1.17 & 0.08 \\
\cite{venugopal} &
$-$22.0 & 3.3 &
1.38 & 0.22 & 1.06 & 0.03 \\
\cite{crystal} &
$-$16.4 & 1.2 &
1.11 & 0.06 & 1.10 & 0.02 \\
\cite{feldmann} &
\multicolumn{2}{l}{$-$15.4} &
\multicolumn{2}{l}{1.26} &
\multicolumn{2}{l}{1.17} \\
\cite{burakovsky} &
$-$21.4 & 1.0 &
1.185 & 0.040 & 1.095 & 0.020 \\
\cite{choi} &
\multicolumn{2}{l}{$-$19.3} &
\multicolumn{2}{l}{1.254} &
\multicolumn{2}{l}{1.127} \\
\cite{escribano} &
$-$18.1 & 1.2 &
1.28 & 0.04 & 1.13 & 0.03 \\
\cite{yeh} &
\multicolumn{2}{l}{$-$16.4} &
\multicolumn{2}{l}{0.99} &
\multicolumn{2}{l}{1.08} \\
\cite{crystal92} &
$-$17.3 & 1.8 &
\multicolumn{2}{l}{---} &
\multicolumn{2}{l}{---} \\
\cite{bramon} &
$-$16.9 & 1.7 &
\multicolumn{2}{l}{---} &
\multicolumn{2}{l}{---} \\
\end{tabular}
\end{table}

\end{document}